\documentclass[fleqn,usenatbib]{mnras}

\usepackage{newtxtext,newtxmath}

\usepackage[T1]{fontenc}
\usepackage{ae,aecompl}
\pdfoutput=1

\usepackage{graphicx}	
\usepackage{amsmath}	
\usepackage{amssymb}	
\usepackage{multirow}

\title[A Gemini/GMOS study of NGC\,3613]{A Gemini/GMOS study of the bright elliptical galaxy NGC\,3613 and its globular cluster system}

\author[De B\'ortoli et al.]{
Bruno J. De B\'ortoli,$^{1,2}$\thanks{E-mails:\,brudebo.444@gmail.com,\,liliaybass@gmail.com, jpceda@gmail.com,\,ennis.ana@gmail.com}
Lilia P. Bassino,$^{1,2}$ Juan P. Caso$^{1,2}$ and Ana I. Ennis$^{1,2}$
\\
$^{1}$Facultad de Ciencias Astron\'omicas y Geof\'isicas de la Universidad Nacional de La Plata, and Instituto de Astrof\'isica de La Plata (CCT \\ La Plata –- CONICET, UNLP), Paseo del Bosque S/N, B1900FWA La Plata, Argentina\\
$^{2}$Consejo Nacional de Investigaciones Cient\'ificas y T\'ecnicas, Godoy Cruz 2290, C1425FQB, Ciudad Aut\'onoma de Buenos Aires, Argentina\\
}

\date{Accepted XXX. Received YYY; in original form ZZZ}

\pubyear{2019}

\begin{document}
\label{firstpage}
\pagerange{\pageref{firstpage}--\pageref{lastpage}}
\maketitle

\begin{abstract}
We present the first photometric study of the globular cluster system (GCS)  of the  E galaxy NGC\,3613 ($M_{\rm V} = -21.5$, d$ \sim  30.1$\,Mpc), as well as the surface photometry of the host galaxy, based on Gemini/GMOS images. Being considered the central galaxy of a group, NGC\,3613 inhabits a low-density environment although its intrinsic brightness is similar to the expected one for galaxies in the centre of clusters. The following characteristics are obtained for this GCS. The colour distribution is bimodal, with metal-poor globular clusters (GCs) getting slightly bluer with increasing radius. The radial and azimuthal projected distributions show that metal-rich GCs are more concentrated towards the host galaxy and trace its light distribution very precisely, while metal-poor GCs present a more extended and uniform distribution. The GC luminosity function helps validate the adopted distance. The estimated total GC population of $N_{\rm tot}= 2075\pm130$  leads to a specific frequency $S_N=5.2\pm0.7$, a value within the expected range for GCSs with host galaxies of similar luminosity. The surface photometry of NGC\,3613 reveals a three-component profile and a noticeable substructure. Finally, a small sample of ultra-compact dwarf (UCD) candidates are identified in the surroundings of the host galaxy.             

\end{abstract}




\section{Introduction}
\label{sec:intro} 
The ages of globular clusters (GCs) usually establish them among the oldest objects in the Universe \citep[e.g.][]{hansen2013,tonini2013}, so they provide important clues about the first phases of galaxy formation. From the observational point of view, GCs present several advantages like being so compact and intrinsically bright that can be observed farther away than one hundred Mpc \citep{harris2014,harris2016,alamo2013}. Moreover, globular cluster systems (GCSs) of early-type massive galaxies contain thousands of GCs, probably as a consequence of a history of numerous mergers \citep[e.g.][]{bassino2008,durrell2014,oldham2016,caso2017}.

It is often assumed that GCs formed at high redshift, in gas-rich discs and within a high-pressure environment \citep{kruijssen2015}. Recent numerical simulations, like the E-MOSAICS Project \citep{pfeffer2018,kruijssen2019}, have presented scenarios that describe the formation, evolution and disruption of the GCs, following their evolution together with that of the host galaxies. These scenarios imply a direct correlation between the formation of GCs and the field stars, in such a way that the properties of GCSs provide constraints to the simulations \citep[e.g.][]{powalka2016b} and, on the other side, a galaxy history can be described based on the study of its GCS. Such interconnections follow clearly from studies of large GC samples, like the ACS Fornax Cluster Survey (ACSFCS) \citep{jordan2007} or the Next Generation Virgo Cluster Survey (NGVS) \citep{ferrarese2012}. 

One of the most common characteristics of GCSs in massive early-type galaxies is the existence of two GC subpopulations, though more complex cases have been pointed out \citep[e.g.][]{caso2013,sesto2016}. These GC subpopulations have been detected through different physical properties: 

\begin{itemize}
\item bimodality in colour, which is interpreted mainly as a difference in metallicity for the {\it bona fide} old GCs, where `blue' and `red' subpopulations identify  those with lower and higher metal content (i.e. metal-poor and metal-rich GCs), respectively  \citep[e.g.][]{usher2012,chiessantos2012,forte2013}.

\item different projected spatial distribution with respect to the host galaxy, with red GCs being generally more concentrated towards the centre of the host galaxies and tracing their surface-brightness profiles, while blue ones present a more extended distribution  \citep[e.g.][]{bassino2006b,forbes2012,durrell2014,escudero2018}.

\item different kinematics, found in the radial velocity and velocity dispersion of the subpopulations. The  kinematics of the red subpopulation is usually akin to that of the host galaxy stars  \citep[e.g.][]{schuberth2010,pota2013,amorisco2019}. According to the numerical simulations by \citet{amorisco2019}, the higher dispersion of blue GCs relative to red ones may be explained by the high contribution of blue clusters to the halo population through minor mergers.\\ 
\end{itemize}

Our current target, NGC\,3613, is an intrinsically bright elliptical galaxy, classified as E6 \citep{devauc1991}. 
We initially adopt a distance d$ \sim 30.1$\,Mpc \citep{tully2013}, based on surface brightness fluctuations, but taking into account that the distances calculated to date have a significant dispersion, as can be seen in NED\footnote{https://ned.ipac.caltech.edu/}.
In particular, one of the aims of this work is to provide a new estimate for this value by means of the turn-over of the globular cluster luminosity function.
Then, the absolute visual magnitude of NGC\,3613 ($M_{\rm V} = -21.5$) corresponds to the range of those of bright massive galaxies located in rich clusters, although it is noticeable that it inhabits an environment of lower density . 

The ATLAS$^{3D}$ project \citep{cappellari2011}, a survey that combines multi-wavelength data and models, includes NGC\,3613 in its sample of 260 early-type galaxies. According to their kinematic analysis \citep{krajnovic2011}, our target is a `regular rotator' (i.e. dominated by ordered rotation) and, based on an estimator of the angular momentum of the stars, it is also classified as a `fast rotator' \citep{emsellem2011}. The local density estimators presented by \citet{cappellari2011b}, place NGC\,3613 in a low-density environment. They also state that fast rotators form a homogeneous category of systems flattened and oblate, with regular velocity fields. One of the last papers of the ATLAS$^{3D}$ project deals with the stellar populations of the early-type galaxy sample \citep{mcdermid2015}, and gives values of the age and metallicity of NGC\,3613, measured within the effective radius, calculated by two methods. Using single stellar population models they obtain: age = $11 \pm 2$\,Gyr and [Z/H] = $ -0.17 \pm 0.05$, and using spectral fitting to derive star formation history, they obtain mass-weighted values of age = $13 \pm 0.7$\,Gyr and [Z/H] = $ -0.13 \pm 0.01$. Under both approaches, our target turns out to be a quite old and metal-poor galaxy.

More recently, \citet{osullivan2017} presented the Complete Local-Volume Groups Sample (CLoGS) that includes 53 optically-selected groups located in the nearby Universe, up to a distance of 80\,Mpc. According to  their selection criterion (i.e. considering the brightest early-type member of the group as the central galaxy), NGC\,3613 is not only member of a group but is also the central galaxy.

As far as we know, the GCS of NGC\,3613 has not been studied before, which is surprising given that the host is such a bright galaxy.
According to the study of \cite{madore2004}, NGC\,3613 belongs to a group consisting of a dozen galaxies. Located at an angular distance of $47$\,arcmin towards the north and with a radial velocity difference of 350\,km sec$^{-1}$, there is a peculiar lenticular (also classified as shell elliptical) galaxy of similar luminosity, NGC\,3610, that is considered as a prototype of a merger remnant of two disc galaxies. The latter galaxy has a very complex surface-brightness distribution with plumes, tails and other structures as a consequence of the tidal disturbances suffered during its evolution \citep{schweizer1992,bassino2017}. 
\cite{madore2004} indicate that both galaxies might belong to the same group, which may then have undergone mergers and tidal-stripping processes. Moreover, the estimated projected distance between them ($ \approx 400$\,kpc assuming they are both at the same distance) lends support to the idea that they may have formed in a common environment.  

Thus, the current analysis of NGC\,3613 and its GCS will not only allow us to characterize the system and confirm its distance, but also look for evidence of possible interactions with other group members, e.g. by detecting spatial enhancements or irregularities in the projected GC azimuthal distribution, substructure in the host surface-brightness distribution, etc. 

This paper is organised as follows. The observations and data reduction are described in Sections \ref{sec:observations} and \ref{sec:phot}, while the results are presented in Section \ref{sec:resu}. In Section \ref{sec:surface} we analyse the surface photometry of the galaxy and present our discussion in  Section \ref{sec:discu}. A summary and conclusions are given in Section \ref{sec:conclu}. 

\section{Observations} 
\label{sec:observations} 

\begin{figure}
	\includegraphics[width=0.5\columnwidth]{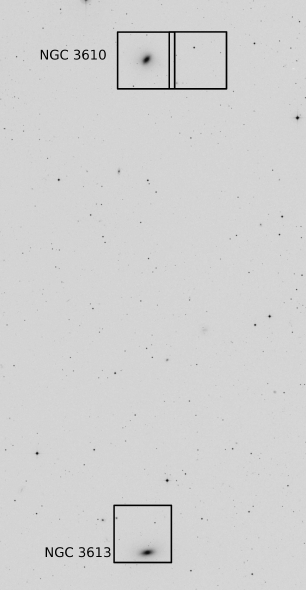}
	\centering
    \caption{ POSS image in $R$-band (field of view FOV: $28 \times 54$\,arcmin). North is up, east to the left. The three GMOS fields (FOV: $5.5 \times 5.5$\,arcmin) observed with our programme are superimposed. The angular distance between both galaxies is approximately 47\,arcmin. Half of the  upper right field is used as a comparison field in the present work.}
    \label{fig:campos}
\end{figure}

The data were obtained with Gemini/GMOS-N during semester 2013A (programme GN2013A-Q-42, PI: J.P. Caso), in nights with photometric quality, and consist of images of the galaxies NGC\,3610 and NGC\,3613 in $g'$, $r'$, and $i'$-bands. Fig.\,\ref{fig:campos} shows the configuration of the observed fields.
The images of NGC\,3610 (one field on the galaxy plus another `adjacent' field) have been used previously to study both the galaxy and its GCS \citep{bassino2017}, while those of NGC\,3613 (one field on the galaxy) are the ones used in the present study to analyse the properties of the GCS of NGC\,3613. In order to estimate the contamination by Galactic stars and background galaxies, we will use half of the `adjacent field' (the one that is further from the centre of NGC\,3610). This field is close in projection and has been taken as part of the same programme. Moreover, the GCS of NGC\,3610 extends up to a galactocentric radius of $\sim 4$\,arcmin, so almost no GCs are expected in this half of the `adjacent field'.


The observing log is presented in Table~\ref{tab:detallesobs}. Four long-exposures were taken for each band. We also note that the $g'$ images were obtained on two different nights.

\begin{table}
	\centering
	\caption{Observation log}
	\label{tab:detallesobs}
	\begin{tabular}{lccr} 
\hline
Filter & N & Date & Exposure time \\
\hline \hline
$g'$ & 2 & 08/03/2013 & 450 sec \\
\hline
$g'$ & 2 & 13/02/2013 & 450 sec \\
\hline
$r'$ & 4 & 13/02/2013 & 210 sec \\  
\hline
$i'$ & 4 & 13/02/2013 & 270 sec \\
\hline
 \end{tabular}
\end{table}

A dithering pattern was used to cover the gaps and remove cosmic rays and bad pixels, as well as a 2x2 binning, resulting in a scale of 0.146 arcsec\,pixel$^{-1}$.

For the data reduction, we used tasks of the \textmd{GEMINI} package (in particular, \textmd{GMOS} package) and \textsc{Iraf} \textmd{DAOPHOT}.

\section{Photometry}
\label{sec:phot}
\subsection{Point source selection and photometry}
In order to improve the detection of GC candidates located near the centre of the galaxy and to remove possible gradients in the surface-brightness distribution, we subtracted the light of the galaxy as much as possible using the task \textmd{FMEDIAN}, applying first a filter that calculates the median value in squares of $200 \times 200$ pixels, and then repeating the procedure with one of $40 \times 40$ pixels to eliminate fluctuations of lower period. With the photometry of the artificial stars (see section \ref{subsec:comp}) we corroborated that this procedure does not modify the results obtained for point sources.

To obtain an initial catalog of point-sources present in the GMOS field, we used the software \textsc{SExtractor} \citep{bertin1996}. 
We ran the software on all $g'$, $r'$ and $i'$ images  
using two filters, one (Gaussian) that is more effective at larger distances from the galaxy, and another (Mexhat) which performs a better fit in highly populated areas such as those near the centre of the galaxy, where candidates for GCs are concentrated. The program generates a catalog for both cases, Mexhat and Gaussian filters. Then, we selected  those objects listed in at least one catalog for  $g'$, $r'$ and $i'$, and with a parameter CLASS\_STAR greater than 0.4 to eliminate extended sources.

We performed PSF photometry with the corresponding tasks of the \textmd{DAOPHOT} package within \textsc{Iraf}. For each filter, a PSF model was obtained with about 20 isolated bright stars, well distributed over the field.
The \textmd{ALLSTAR} task also gave us statistical parameters ($\chi$ and sharpness). By means of these parameters, a new improved point-source catalog was obtained. 

Finally, aperture corrections were estimated using the same objects as those used to obtain the respective PSFs. 

\subsection{Photometric calibration}
As part of the Gemini programme, a standard star field from the list of \cite{smith2002} was observed, and reduced in the previous study of NGC\,3610 by \cite{bassino2017}. To obtain magnitudes in the standard system, the calibration equation for each filter is:
\begin{equation}
m_{std}=ZP+m_{inst}-K_{MK}\,(X-1.0)
\end{equation}
where $m_{std}$ and $m_{inst}$ are the standard and instrumental magnitudes, respectively, $ZP$ is the photometric zero-point, $K_{MK}$ the mean atmospheric extinction at Mauna Kea (obtained from the Gemini Observatory Web Page \footnote{http://www.gemini.edu/sciops/instruments/gmos/calibration}), and $X$ the airmass. In the present work, the same calibration equations as those obtained by \cite{bassino2017} were applied.

Finally, we applied the corrections by Galactic extinction obtained from NED, which were calculated by \cite{schlafly2011}.

\subsection{Completeness estimation}
\label{subsec:comp}
\begin{figure}
	\includegraphics[width=\columnwidth]{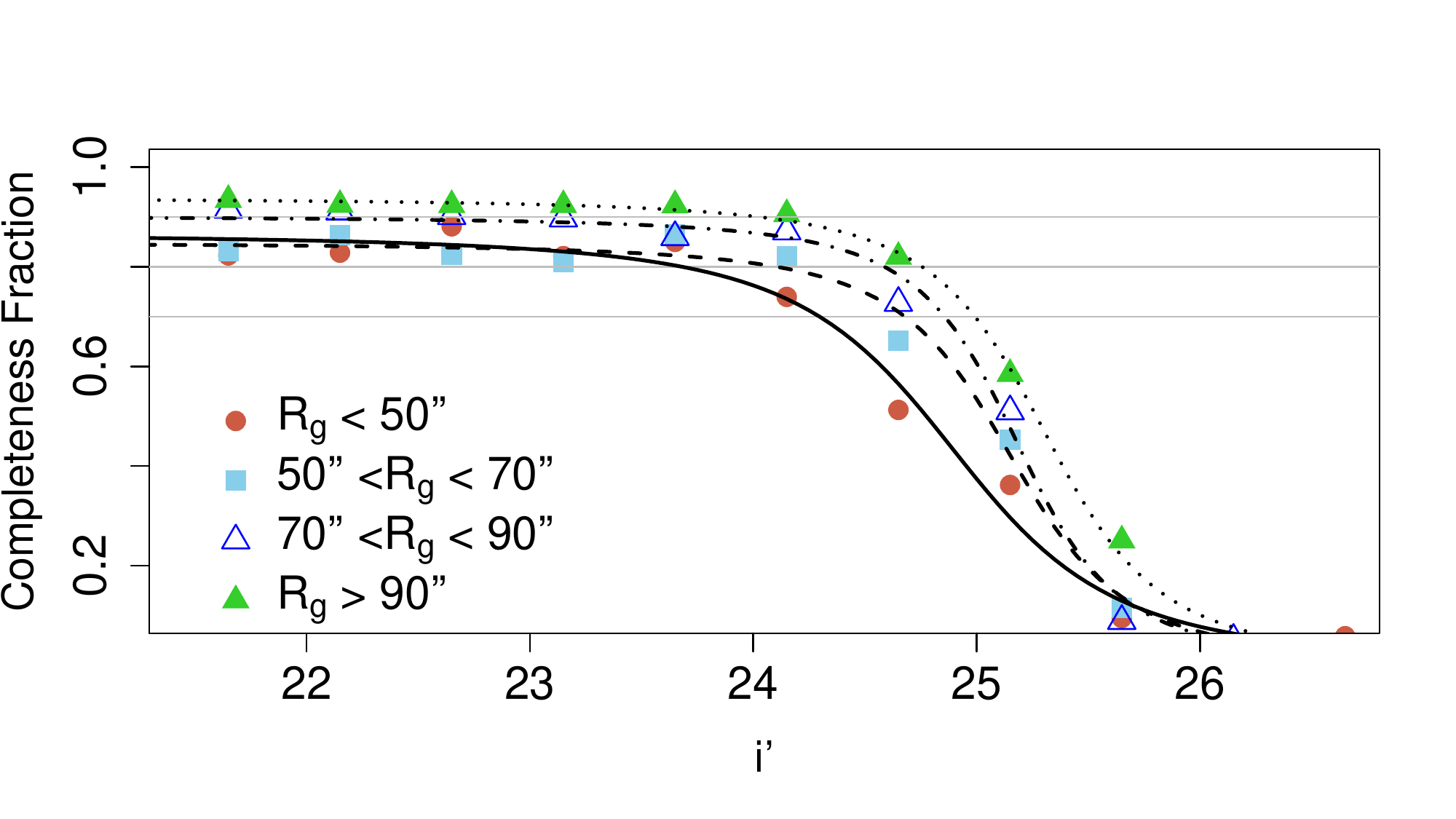}
    \caption{Completeness curves for the science field as a function of $i'_0$. Different line types represent the fits for the denoted $R_g$ ranges. Horizontal lines indicate 70, 80 and 90 per cent completeness.}
    \label{fig:completitudciencia}
\end{figure}

\begin{figure}
	\includegraphics[width=\columnwidth]{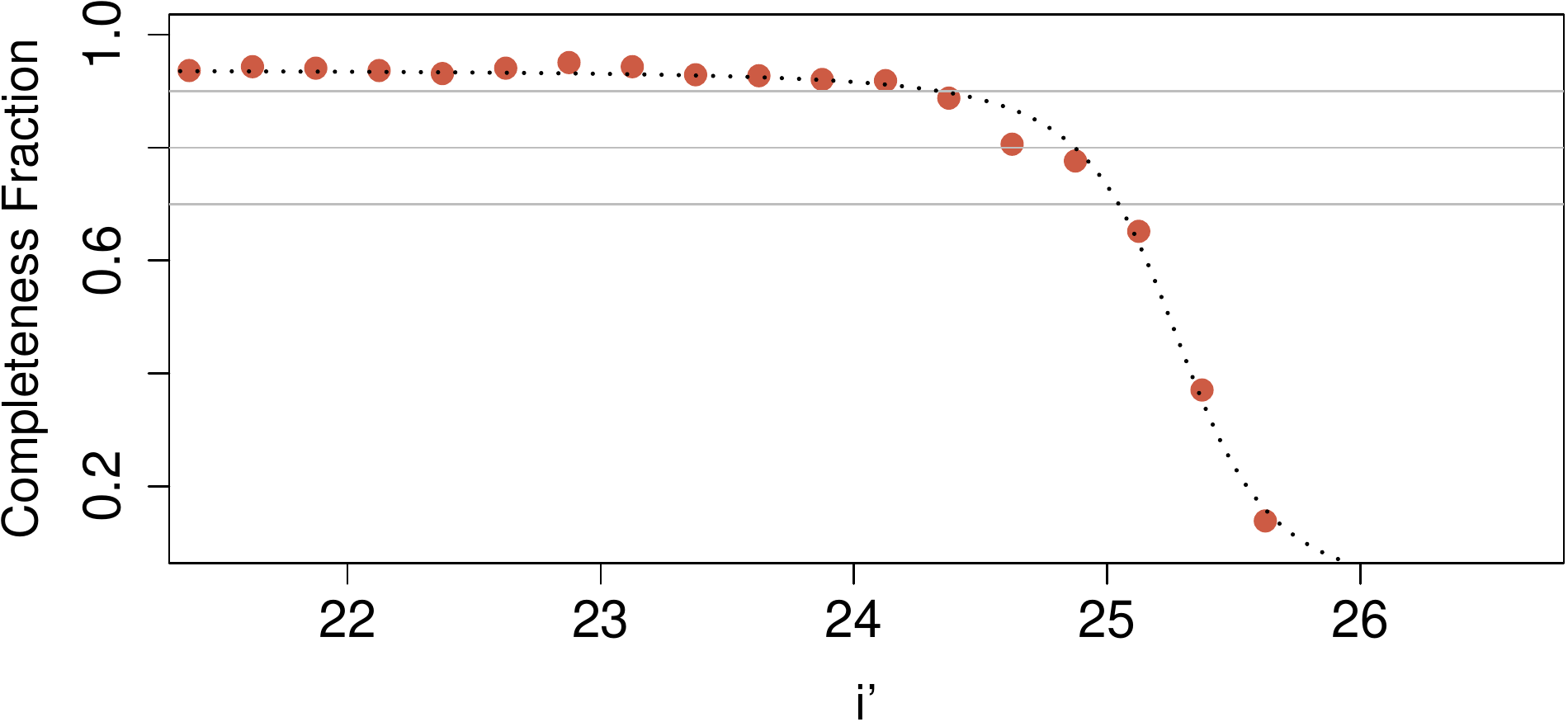}
    \caption{Completeness curve for the comparison field as a function of $i'_0$. Dotted line represents the fit. Horizontal lines as in  Fig.\,\ref{fig:completitudciencia}.}
    \label{fig:completitudcomparacion}
\end{figure}

In order to estimate the photometric completeness for our fields, first we added to the images 250 artificial stars, uniformly distributed, covering a magnitude range 21.5$ \leq i'_0 \leq $ 27 and the expected colour range for GCs. We repeated this procedure 40 times, achieving a sample of 10\,000 artificial stars in each image. 
Then, we performed the detection and photometry in the same way as in the original science images. The process was carried out for four
ranges of galactocentric radii. In addition, it was repeated for the comparison field in order to estimate the contamination corrected by completeness. The resulting completeness curves are shown in Figs.\,\ref{fig:completitudciencia} and \ref{fig:completitudcomparacion}. 

The fitted function \citep{harris2009b} is:
\begin{equation}
f(i'_0) = \beta \, \left( 1 - \frac{\alpha(i'_0-m_0)}{\sqrt{1+\alpha^2(i'_0-m_0)^2}}\right) 
\end{equation}
where $\alpha$, $\beta$ and $m_0$ are the free parameters.

Hereafter, the limit $i'_0 <25$ is used to guarantee an acceptable completeness in both, the science and comparison fields.

\section{Results}
\label{sec:resu}
\subsection{Selection of GC candidates}
The GC candidates will be selected among the point-sources, according to certain brightness and colour ranges. On the one hand, the faint magnitude limit was determined in the previous Section according to the adopted completeness for the science images (i.e. $i'_0 \sim 25$). On the other hand, the bright magnitude limit will be taken as the estimated limiting magnitude that separates Ultra Compact Dwarf (UCD) and bright GC candidates. Adopting as such limit the $M_I$ magnitude derived from \citet{mieske2006} (i.e. $M_I = -12$), using the transformation equations given by \citet{faifer2011} and the adopted distance for NGC\,3613, we calculate the bright magnitude limit as $ i'_0 = 20.8 $.

\begin{figure}
	\includegraphics[width=\columnwidth]{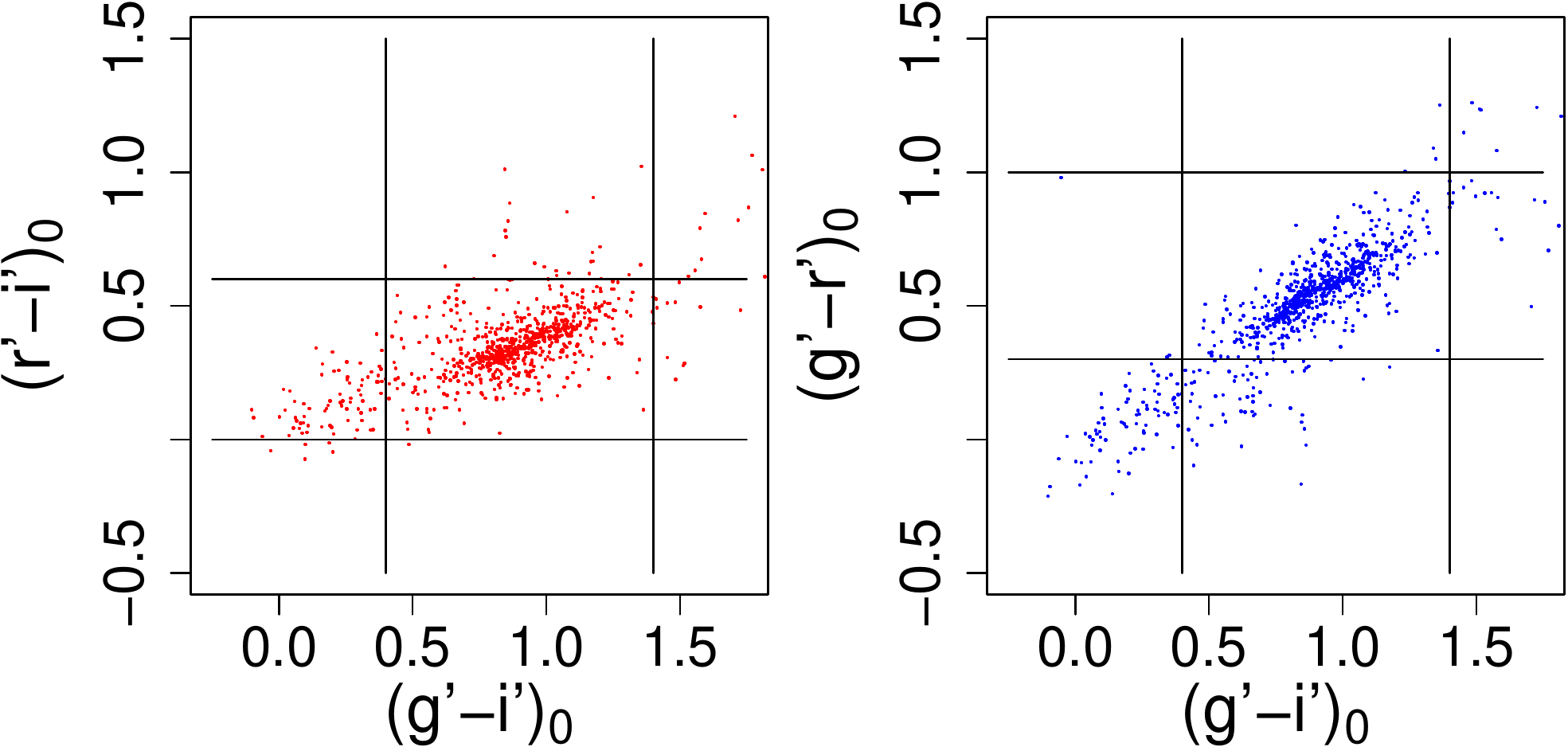}
    \caption{Colour-colour diagrams for the selected  point-sources. Solid lines indicate the typical colour limits for GC candidates.}
    \label{fig:dcc}
\end{figure}

In regards to the colour range, Fig.\,\ref{fig:dcc} shows the colour-colour diagrams, $(r'-i')_0$ versus $(g'-i')_0$ and $(g'-r')_0$ versus $(g'-i')_0$ for the selected point-sources. The use of colour-colour diagrams to select GC candidates has been thoroughly explained by \citet{faifer2011}. This method has proved to result in a clean selection of GCs, with only a small fraction of contaminants, when spectroscopic observations to confirm membership are available \citep[e.g.][and references therein]{norris2008,norris2012}
Accordingly, there are well-defined sequences in these diagrams that are indicated by the solid lines. We then select as GC candidates those in the colour ranges $ 0.4 <(g'-i') _ 0 <1.4 $, $ 0.0 <(r'-i') _ 0 <0.6 $ and $ 0.3 <(g'-r ') _ 0 <1 $  \cite[and references therein]{caso2015,escudero2015}.

\begin{figure}
	\includegraphics[width=\columnwidth]{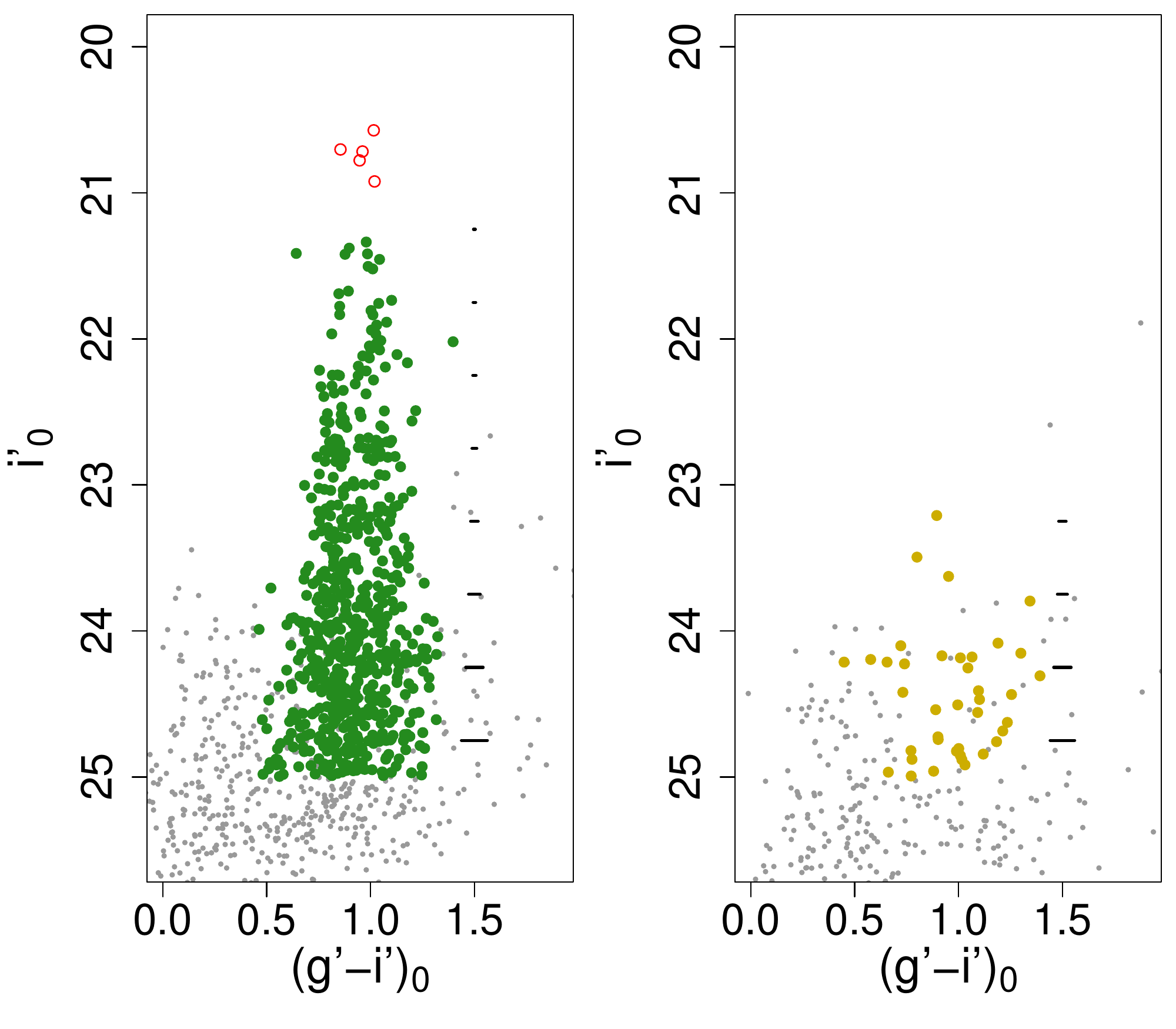}
    \caption{Colour-magnitude diagram for the science field (left panel) and the comparison field (right panel). The highlighted points represent those specific objects that fulfill the criteria adopted for selecting GC candidates. The open circles indicate UCD candidates (see Discussion). The colour errors for different magnitudes are displayed on the right side of both diagrams.}
    \label{fig:dcm}
\end{figure}

Finally, Fig.\,\ref{fig:dcm} shows the colour-magnitude diagram $ i'_0$ versus $(g'-i ') _ 0 $ for the science field (left panel) and for the comparison field (right panel). The locus of the {\it bona-fide} GC candidates appears clearly on the science field, even the two subpopulations can be distinguished at first sight. In the comparison field, contaminants that fulfill the same criteria as GC candidates are present only for  $ i'_0 > 23 $, with a total of 4.9\,objects/arcmin${^2}$.   

\subsection{Colour distribution}
\begin{figure}
	\includegraphics[width=\columnwidth]{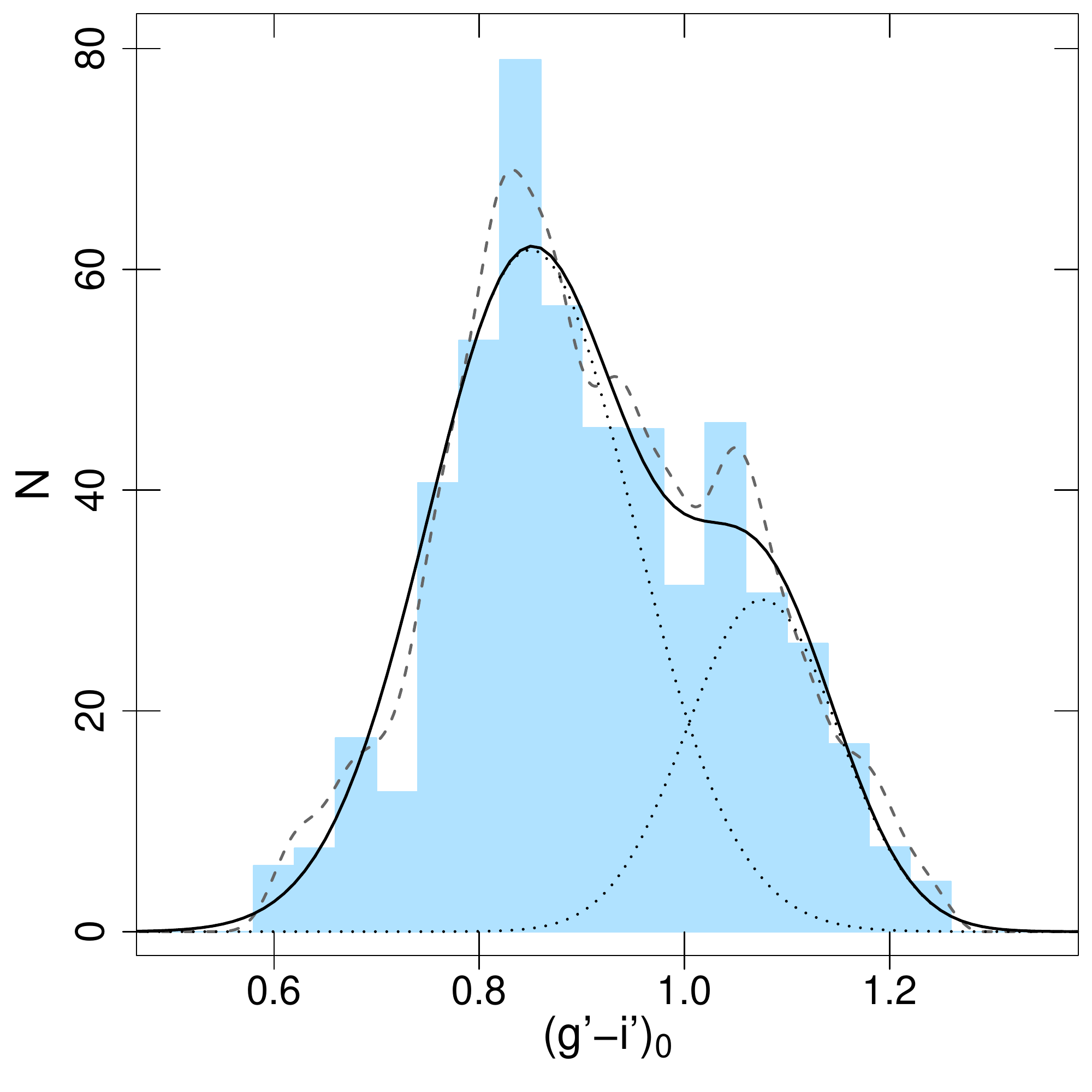}
    \caption{Colour-distribution for GC candidates. Dashed and dotted lines show a smoothed histogram (with a 0.5$\sigma$ Gaussian kernel)} and the two fitted Gaussians, respectively. Solid line shows the result of adding these Gaussians.
    \label{fig:dcol}
\end{figure}

Fig.\,\ref{fig:dcol} shows the $(g'-i') _ 0$ colour distribution for all GC candidates, using a bin width of 0.04 mag. A smoothed histogram (with a 0.5$\sigma$ Gaussian kernel) is also shown with dashed-lines. We note that for this analysis, the central zone of the galaxy is excluded due to saturation. 

In order to analyse whether the global colour distribution can be represented as the sum of two Gaussian models, we used the Gaussian Mixture Modeling test (GMM, \citealt{muratov2010}). By means of the GMM test, we fitted two Gaussians to the sample, obtaining mean value, dispersion, and fraction for each subpopulation, i.e. metal-poor (`blue') and metal-rich (`red') GC candidates. The test also gives two statistical parameters, DD and the kurtosis of the input distribution. The DD parameter is a measure of the separation between the peaks of the two Gaussians, calculated as:
\begin{equation}
DD=\frac{|\mu_1 - \mu_2|}{(\sigma_1^2+\sigma_2^2)^{1/2}}
\end{equation}
where $\mu_1$ and $\mu_2$ are the mean values and $\sigma_1$ and $\sigma_2$ the dispersions of the fitted Gaussians.
A bimodal distribution is acceptable when DD > 2, while the kurtosis is very likely negative in such a case.

In order to run GMM on contamination-free samples, we proceeded as follows. The expected number of contaminants, $ N_c $, was calculated for each region, taking into account the ratio between the areas covered by the sub-sample and the comparison field. 
Due to the fact that the regions in which the sample was divided present a smaller area than that corresponding to the comparison field, we proceeded to randomly select $ N_c $ objects from the comparison field, to then subtract from the science sample those that present more similar colours to each of them.
This random selection can introduce some statistical noise. To minimize this effect, the procedure was repeated 25 times and the results were averaged to obtain the final parameters of each fitted Gaussian.

The results are listed in Table~\ref{tab:paramgmm}, where it can be seen that for the whole sample it is acceptable to consider a bimodal distribution.

\begin{figure}
	\includegraphics[width=\columnwidth]{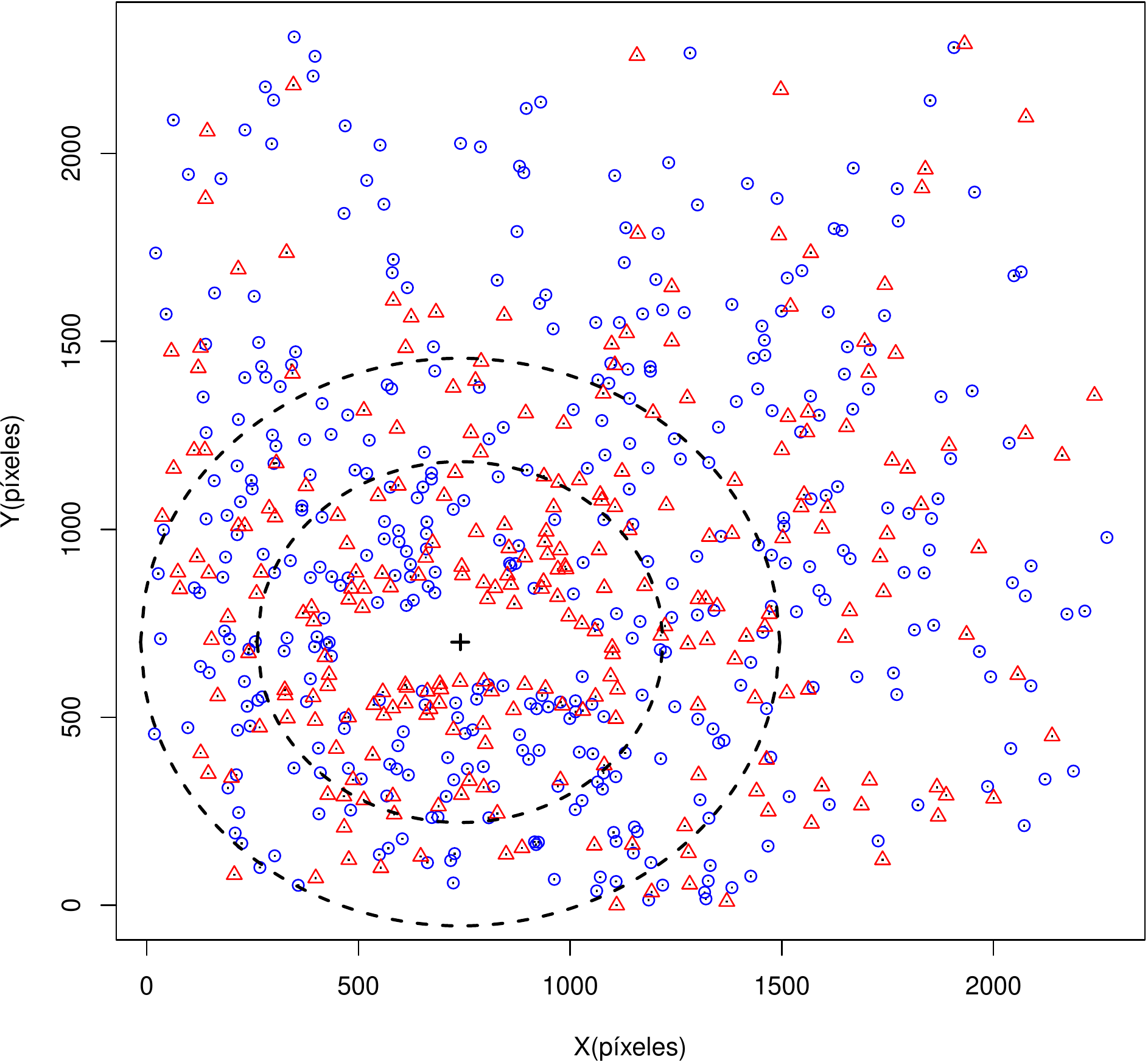}
    \caption{ Projected spatial distribution of blue and red GC candidates, indicated with blue circles and red triangles, respectively.
    The dashed lines show the three radial ranges used to study the colour distribution. The galaxy centre is marked with a cross.
    }
    \label{fig:espadcol}
\end{figure}

\begin{figure*}
	\includegraphics[width=1.5\columnwidth]{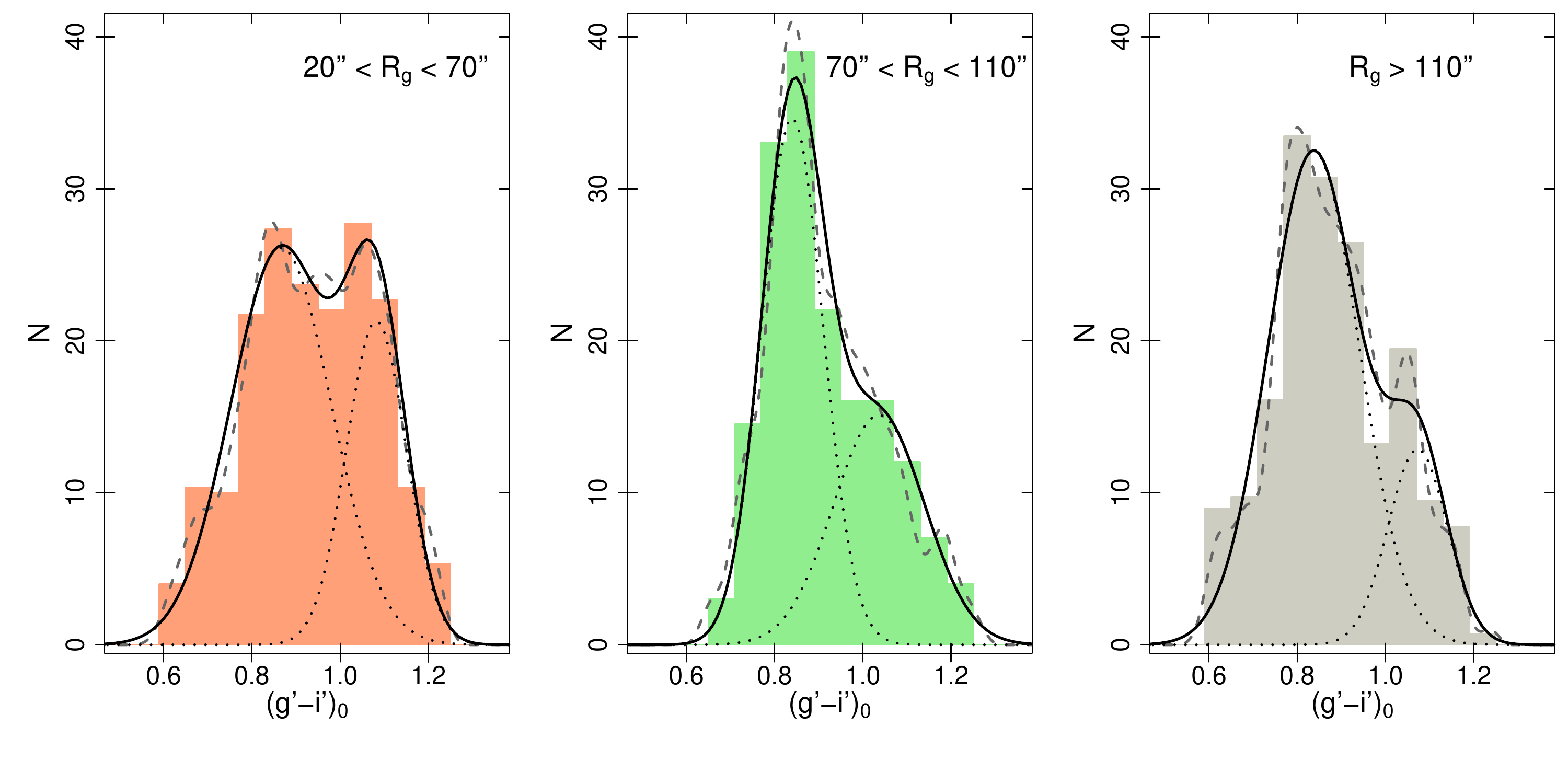}
    \caption{Colour distribution for GC candidates, for three different radial regimes. Dashed, dotted, and solid lines as in Fig.\,\ref{fig:dcol}.}
    \label{fig:dcol2}
\end{figure*}

We also performed this analysis for three concentric regions. We separated them according to the following galactocentric radii ($ R_g $): 20 $< R_g <$  70\,arcsec ,  70 $ <R_g < $ 110\,arcsec,  and $ R_g > $ 110\,arcsec (see Fig.\,\ref{fig:espadcol}), using a bin width of 0.06 mag. Fig.\,\ref{fig:dcol2} depicts the three colour distributions and Table~\ref{tab:paramgmm} shows the  corresponding results of the GMM test. According to the DD parameters and kurtosis obtained, it is also acceptable to consider bimodal distributions for the subsamples in the three concentric regions. As the $ f_ {red} $ parameter indicates, the blue subpopulation dominates clearly in all galactocentric ranges, unlike other bright elliptical galaxies where in the innermost region the weight of both subpopulations is similar \citep[e.g.][]{caso2019}.

\begin{table*}
	\centering
    \caption{
    Parameters obtained by fitting two Gaussians with the GMM test, for different radial ranges. $ \mu_1 $, $ \sigma_1 $,   $ \mu_2 $, $ \sigma_2 $ correspond to the mean value and dispersion in $ (g'-i ') _ 0 $ for the blue and red subpopulations, respectively. DD and kurt (kurstosis) are obtained with GMM. The fraction of red GCs $f_red$, is depicted in the last column.
    }
    \label{tab:paramgmm}
\begin{center}
\begin{tabular}{|c|c|c|c|c|c|c|c|}
\hline
 Region & $\mu_1$ & $\sigma_1$ & $\mu_2$ & $\sigma_2$ & DD & kurt & $f_{red}$ \\
\hline \hline
 Total & $0.850\pm0.003$ & $0.100\pm0.002$ & $1.076\pm0.006$ & $0.074\pm0.005$ & $2.538\pm0.085$ & -0.573 & $0.265\pm0.020$\\
\hline
$20" < R_g < 70"$ & $0.866\pm0.006$ & $0.114\pm0.002$ & $1.082\pm0.005$ & $0.068\pm0.003$  & $2.294\pm0.058$ & -0.776 & $0.327\pm0.024$ \\
\hline
$70" < R_g < 110"$ & $0.840\pm0.006$ & $0.070\pm0.008$ & $1.037\pm0.029$ & $0.104\pm0.010$ & $2.149\pm0.359$ & -0.226 & $0.393\pm0.097$ \\
\hline
$R_g > 110"$  & $0.837\pm0.009$ & $0.103\pm0.003$ & $1.073\pm0.019$ & $0.067\pm0.016$  & $2.634\pm0.251$ & -0.486 & $0.205\pm0.042$\\
\hline
 \end{tabular}
 \end{center}
\end{table*}

As can be noticed from Table~\ref{tab:paramgmm}, mean $(g - i)'_0$ colours of blue and red subpopulations remain approximately at similar values for the three subsamples and for the total population, except that the blue peak gets bluer with increasing radius (we will come back to this in the Discussion) and the red peak of the intermediate region is bluer than the rest, though the latter also has the largest error. Globally, these mean values mostly agree with those found in other studies of GCSs in the same photometric system, that is $\mu \approx 0.85$ and $\mu \approx 1.07$ for the blue and red peaks, respectively \citep[e.g.][and references therein]{harris2009b,forbes2011}.
Moreover, the fraction of metal-rich clusters in the inner and intermediate regions is larger than in the outermost one, which is in agreement with the idea that this red subpopulation is more concentrated towards the host galaxy and thus, closely related to its stellar component.

\subsection{Blue-tilt}
In the colour-magnitude diagram depicted in Fig.\,\ref{fig:dcm}, it can be clearly seen that as we consider brighter blue GC candidates, they get redder. This behaviour has been generally called `blue-tilt' and, in our case, it extends over the whole luminosity range. Also, some authors refer to it as a `mass-metallicity relation (MMR)' \citep[e.g.][]{harris2006}, applied to this colour-luminosity trend followed by the metal-poor GCs in many bright galaxies, but not all of them. 

\begin{figure}
	\includegraphics[width=0.8\columnwidth]{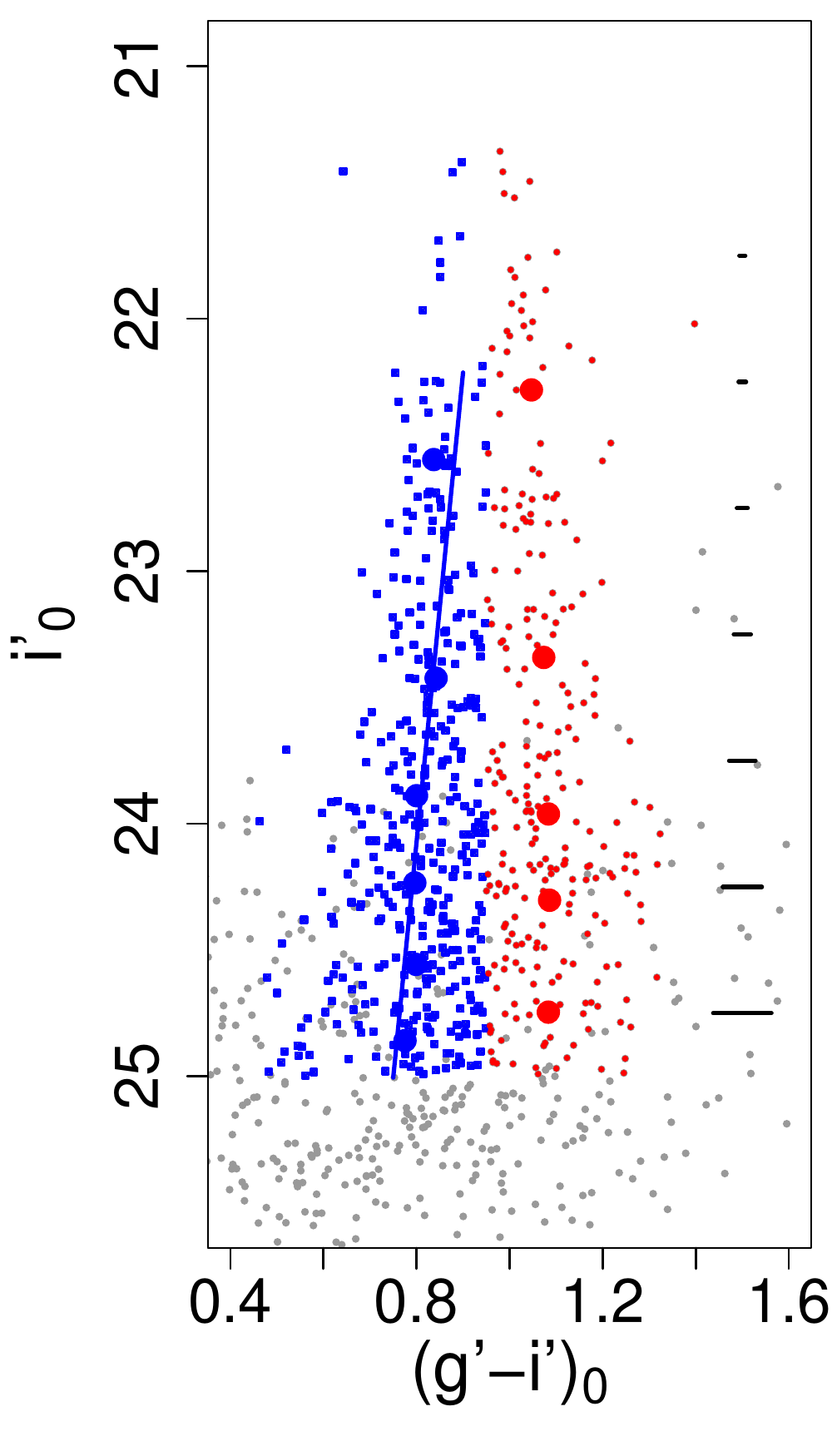}
    \caption{Colour-magnitude diagram for point sources (small dots). Metal-poor (blue) and metal-rich (red) GC candidates are shown with squares and filled circles respectively. Large dots represent colour averages of different subsamples, sorted in magnitude,  with equal number of candidates. 
    }
    \label{fig:bluetilt}
\end{figure}

In order to characterize the blue-tilt, Fig.\,\ref{fig:bluetilt} shows the colour-magnitude diagram, differentiating the red and blue GC candidates by taking $(g-i)'_0=0.95$ \citep{faifer2011} as limiting colour between both subpopulations. In addition, the large dots represent the mean colour of different adjacent subsamples in each subpopulation, each subsample with equal number of GC candidates (50 for the red candidates and 65 for the blue ones). 
It can be seen that in the case of the blue GC candidates, a correlation between colour and magnitude is present, as mean colours are tilted towards the red as we consider brighter GCs. By means of a linear least-squares fit of those mean blue colours we obtained a slope of $d(g'-i')_0/di'_0=-0.053\pm0.015$ (the result of a chi-square test indicates that the fit represents the distribution with a 90 per cent of confidence.). Thus, it is in agreement within uncertainties to that obtained, in the same photometric system, by \citet{wehner2008} ($d(g'-i')_0/di'_0=-0.044\pm0.011$) for NGC\,3311, the central galaxy of the Hydra cluster, and slightly larger than the one obtained by \citet{escudero2015}($d(g'- i')_0/di'_0=-0.026\pm0.007$)  for a bright lenticular, NGC\,6861.

\subsection{Projected spatial and radial distributions}
\begin{figure*}
	\includegraphics[width=1.8\columnwidth]{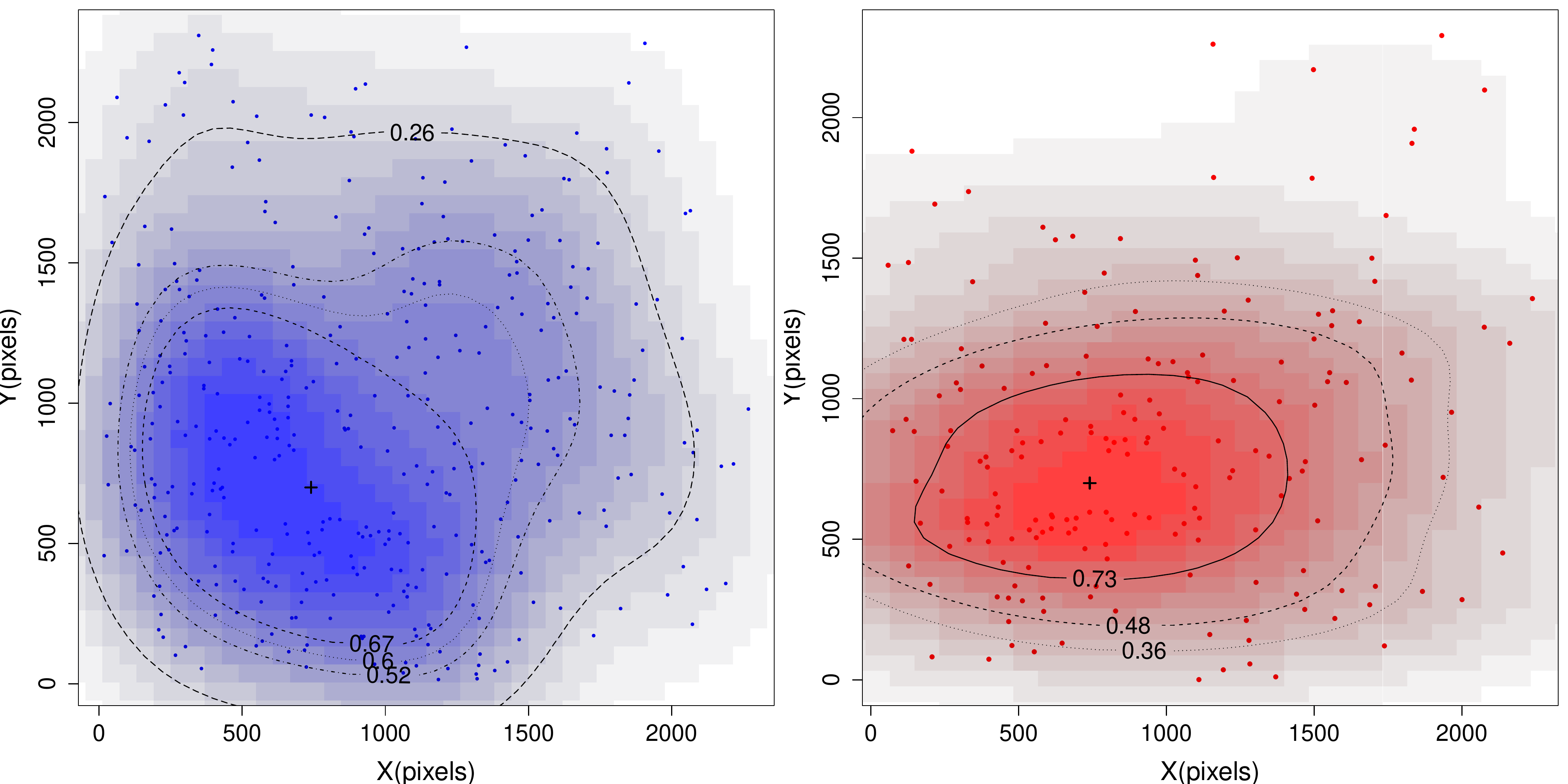}
    \caption{Projected spatial distribution (dots and smoothed distribution) for the blue (left panel) and red (right panel) GC subpopulations.
    The lines show contours of equal projected number density (darker colours represent higher densities and vice versa). The numbers indicate the value with respect to the maximum.
    The galaxy centre is marked with a cross.
    }
    \label{fig:espa}
\end{figure*}

Fig.\,\ref{fig:espa} shows the projected spatial distribution of the GC candidates surrounding the galaxy NGC\,3613. It is divided into blue and red GC subpopulations, according to the adopted colour limit, $(g - i)'_0 = 0.95$. The corresponding projected density is superimposed as a smoothed distribution as well as a few contours of constant numerical  density. 

As already indicated by the decreasing fraction of red GCs with galactocentric distance, it is clear from Fig.\,\ref{fig:espa} that the red GC subpopulation is more concentrated towards the centre of the galaxy, while the blue subpopulation is more extended and evenly distributed in an approximately circular distribution. The contours of the red GCs are elliptical, with the major axis oriented in a similar direction as the host galaxy starlight. 

\begin{figure}
	\includegraphics[width=\columnwidth]{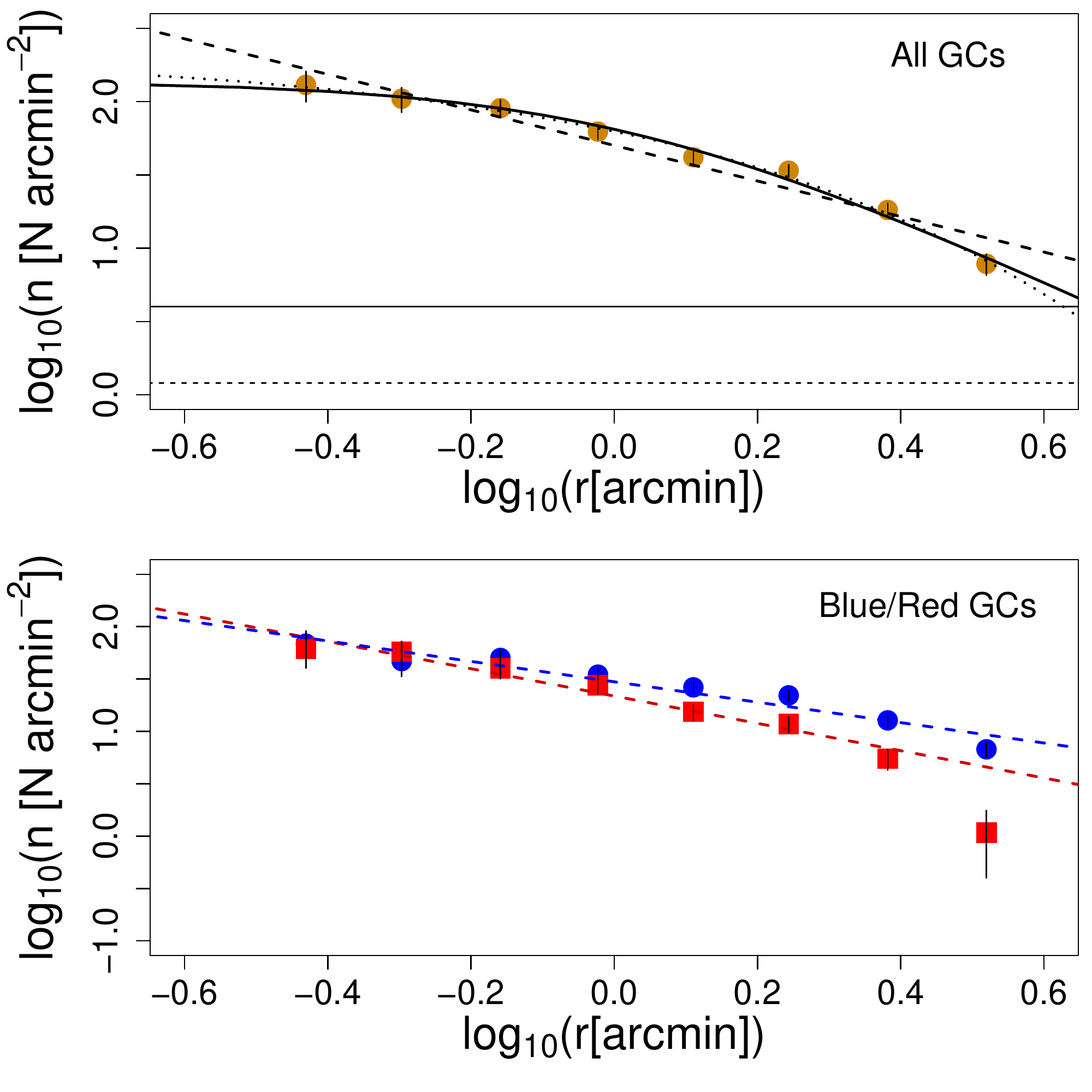}
    \caption{Projected radial distribution, corrected by background and completeness. Upper panel: for the total GC population, lower panel: for the blue and red GC subpopulations (circles and squares, respectively).Solid, dotted, and dashed lines show fits using a Hubble profile, a Sérsic model, and a power-law, respectively.}
    The solid horizontal line represents the value of the background level and the dotted horizontal line a 30 percent of the background level, which is used to estimate the extension of the GCS.  
    \label{fig:drad}
\end{figure}

The projected radial distributions for all GC candidates and for both  subpopulations, corrected by contamination and by completeness, are presented in Fig.\,\ref{fig:drad}. 
All the radial profiles were fitted with power-laws to calculate the respective slopes. Due to saturation at the galaxy centre, the fits were performed for r > 0.35\,arcmin. 
According to the power-law, 
\begin{equation}
log_{10}(N)=d+e\,log_{10}(r)
\end{equation}
r is the galactocentric radius and d, e are the fitted coefficients.
The corresponding results are depicted in Table~\ref{tab:paramrad}. As can be seen in Fig.\,\ref{fig:drad}, the power-law provides good fits for the blue and red subpopulations, excluding from the fit of the latter subpopulation the furthermost point. However, the power-law fit is not as good for the case of the whole sample. 

Then, a modified Hubble distribution \citep{binney1987} was also fitted to the whole sample profile, within the same radial range, to take into account the evident change of slope present in the profile. In previous works we have obtained good fits this way \citep[e.g.][]{bassino2017,caso2017}. By means of the Hubble profile, 

\begin{equation}
n(r) = a \left( 1+ \left(\frac{r}{r_0} \right)^2 \right)^b
\end{equation}
where $r$ is the galactocentric radius and $a$, $b$, $r_0$ are the fitted coefficients, we obtained the following values $a=137\pm 15$\,$N\,arcmin^{-2}$,  $~r_0=1.04\pm0.24$\,arcmin and $b=-1.15\pm0.19$. This fit is better than the one obtained with a power-law, particularly for the innermost points where the destruction of GCs \citep{kruijssen2012,kruijssen2015} must have affected the profile. As a third option, we fitted a Sérsic model \citep{sersic1968} to the whole GC sample, that resulted quite similar to that of the Hubble profile and gave an effective radius $R_{\rm eff} = 1.97 \pm 0.16$\,arcmin (17\,kpc), for the total projected GC distribution (Fig.\,\ref{fig:drad}, upper panel). This value is slightly larger than those obtained by \citet{usher2013} for NGC\,4278 (12.7\,kpc), and \citet{kartha2014} for NGC\,720 and NGC\,2768 (13.4\,kpc and 10.5\,kpc, respectively), all E galaxies with similar luminosity than NGC\,3613.

We assume that the total extension of the GCS is reached at the radius where the background-corrected density, corresponding to the Hubble profile, is equal to 30 per cent of the background level. Such a criterion 
was first used by \citet{bassino2006b} in a wide-field study of the GCS of NGC\,1399, based on three MOSAIC\,II (CTIO) images (FOV: $36 \times 36$\,arcmin each). The galactocentric radius corresponding to 30 per cent of the background was the largest distance from the host galaxy where GCs and the background could be separated, being the density distribution flat further out.
If we consider this limit, which has also been adopted in subsequent works  \citep[e.g.][]{caso2013,bassino2017}, the GCS of NGC\,3613 exceeds the FOV of our images. Thus, we obtain an extension of $r=8.1$\,arcmin, that is $r=70$\,kpc.

\begin{table}
\begin{center}
\begin{tabular}{|c|c|c|c|}
\hline
    & All &  Blue  & Red\\
\hline \hline
d & $1.70\pm0.03$ & $1.47\pm0.03$ & $1.33\pm0.03$\\
\hline
e & $-1.21\pm0.12$ & $-0.97\pm0.10$ & $-1.30\pm0.11$\\
\hline
 \end{tabular}
      \caption{Coefficients of the power-law fitted to the radial profiles for all, blue, and red GC candidates.} 
         \label{tab:paramrad}
 \end{center}
\end{table}

\subsection{Azimuthal distribution}
\begin{figure}
	\includegraphics[width=\columnwidth]{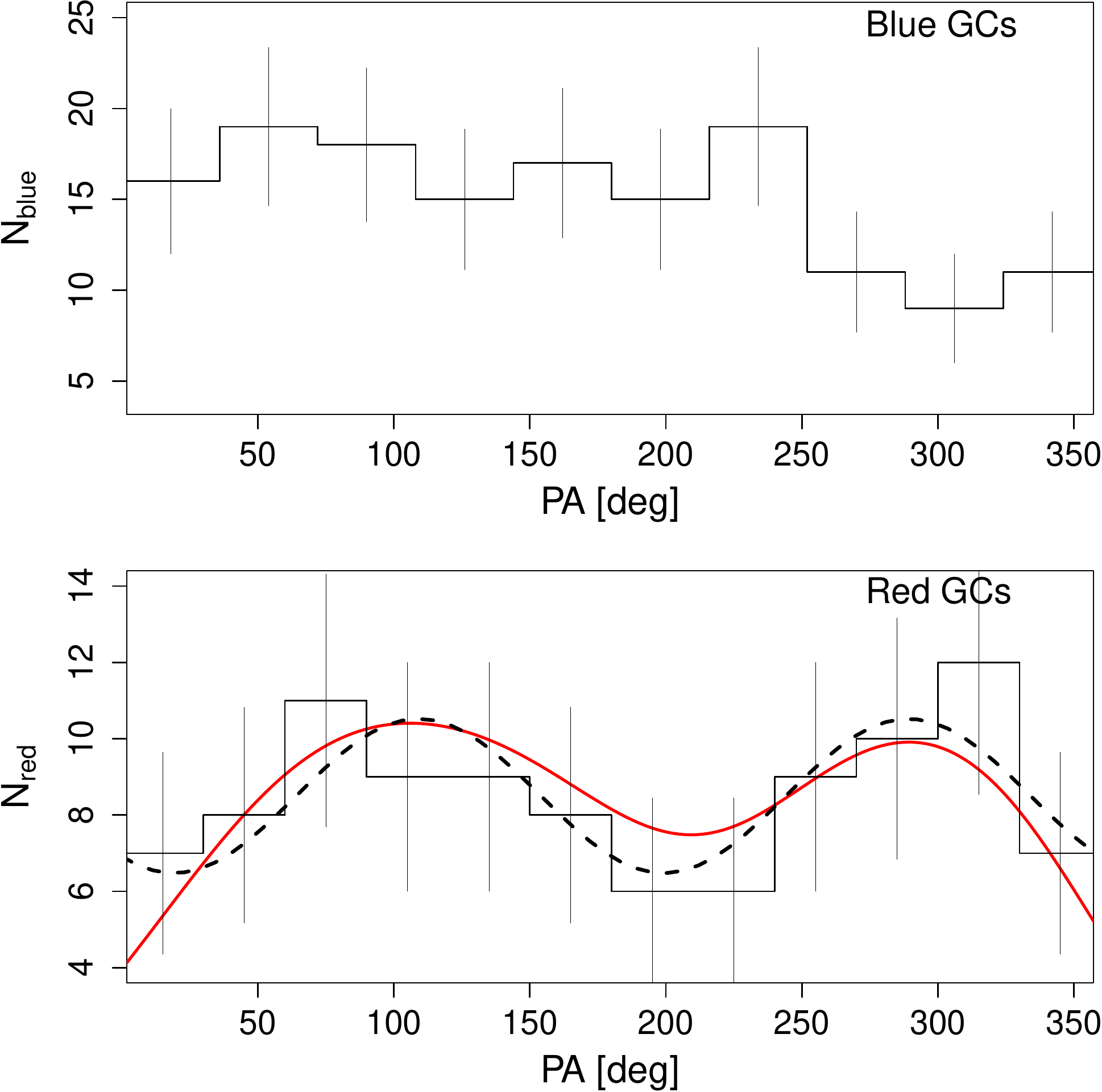}
    \caption{Projected azimuthal distribution for the blue (upper panel) and red (lower panel) GC subpopulations.
    The dashed and solid lines show the fit of a sinusiodal    function (see text) and the smoothed histogram respectively. 
    }
    \label{fig:dacim}
\end{figure}

Fig.\,\ref{fig:dacim} shows how the GC subpopulations are distributed with respect to the position angle (PA), which is measured from north to east with vertex at the galaxy centre. Such distributions were estimated considering an annulus defined by the largest possible outer radius so as the whole annulus was contained within the FOV, i.e. 48 $ < R_g < $ 102\,arcsec. It was divided into angular sections of $\approx 30^{\circ}$ and the GC number density was calculated for each bin.
It can be seen that the blue GCs do not show any particular behavior, as it is basically a rather uniform distribution, except for a slight drop at PA $\sim 300^{\circ}$. On the other hand, the red GCs show a sinusoidal behavior, with two clear over-densities at PA that differ approximately by $180^{\circ}$. As expected, the position of these over-densities agrees with what is obtained from the contours of constant density at the GC projected spatial distribution (see Fig.\,\ref{fig:espa}), defining the same direction as  the ellipse major axis.

In order to fit the red GC distribution, we used the sinusoidal function:
\begin{equation}
N_{red}=A+B\,sin(2\,PA+\phi)
\end{equation}
where $N_{red}$ is the density of red clusters, PA is the position angle, A is the offset of the symmetry axis, B is the amplitude and $\phi/2$ is the phase shift.

The parameters resulting from the fit are $A=8.50\pm0.36$, $B=-2.02\pm0.51$ and $\phi=52^{\circ}\pm14^{\circ}$. According to them, the PA of the maximum. i.e. the first over-density, is $\sim 109^{\circ}$.

We also calculated the ellipticity of the projected distribution of the red GCs by means of the  expression proposed by \cite{dirsch2003} and obtained a value of $\epsilon=0.37$.

We note that this analysis of the azimuthal distribution applies to just a fraction of GC candidates, those located within the annulus defined above, while the rest of the GC population is not included. In addition, as the photometry of objects that are close to the borders of the image is usually not very accurate, the outer radius of the annulus was reduced.

\subsection{Luminosity function and GC population}
\begin{figure}
	\includegraphics[width=\columnwidth]{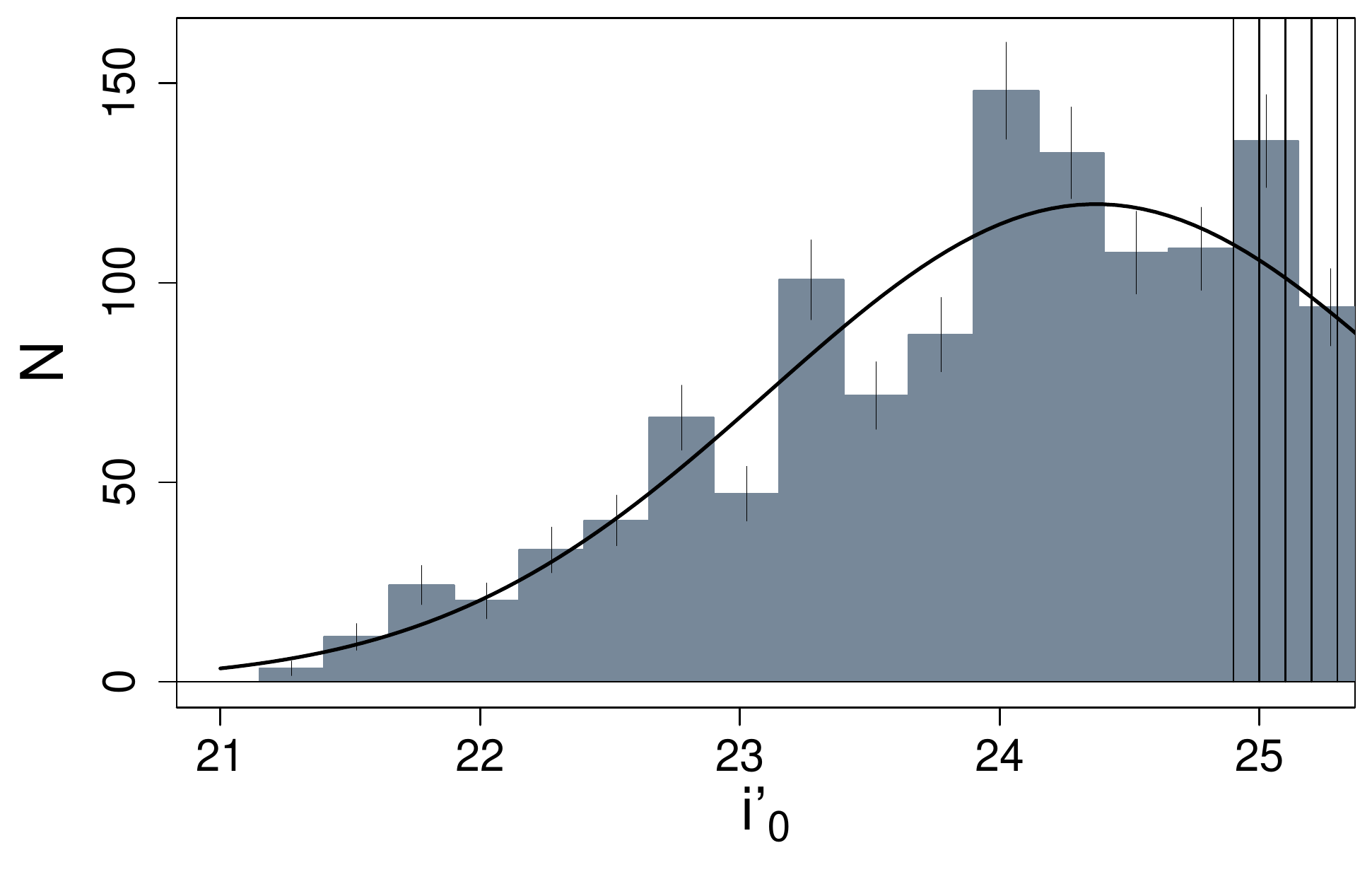}
    \caption{Luminosity function for GC candidates in i'-band, background and completeness-corrected. Errors are estimated as Poisson uncertainties. Solid line shows a Gaussian fit. 
    Vertical lines show the magnitude range that was not considered for the fits. 
    }
    \label{fig:lumin}
\end{figure}

Fig.\,\ref{fig:lumin} shows the background and completeness corrected globular cluster luminosity function (GCLF), using a binwidth of 0.25\,mag. 
Two Gaussians were fitted to the GC candidates with i' $\leq$\,24.9, excluding fainter ones due to declining completeness. One fit was performed leaving all the parameters free (solid line) and the other one using a fixed mean (turn-over), which was calculated with the adopted distance modulus and an universal absolute visual magnitude $M_{\rm V_0} = -7.4$, taken from \citet{richtler2003}. Afterwards, we converted $V_0$  to $i'_0$ using the transformations given by \cite{bassino2017}. There are no notable differences between the two options. Therefore, from here on we will consider the results of the Gaussian fitted with all parameters free. We obtained a turn-over $i'_0 = 24.37\pm0.25$ with a dispersion of $1.26\pm0.20$, that corresponds to a distance modulus $(m-M)=32.37\pm0.2$. This value is in agreement within uncertainties  with the distance modulus $(m-M)=32.39 \pm0.14$ given by \citet{tully2013}, which is based on surface brightness fluctuations.

In order to calculate the GC population, we integrated the Hubble law fitted to the radial distribution, assuming that a background-corrected density of 30 per cent of the background sets the limit of the system (see Section\,4.4). 
Afterwards, we applied another correction to take into account that, according to the GCLF, this first result corresponds to only GCs brighter than $ i'_0 = 24.9 $ and we want to consider the whole population. 
Finally, we obtained a total GC population of $N_{\rm tot}= 2075\pm130$ members.

The specific frequency $S_N$ is defined as the number of GCs per unit $M_{\rm V}$ of host galaxy luminosity \citep{harris1981}, which was considered to be closely linked to the formation efficiency of GCs \citep{mclaughlin1999}. We obtained a value $ S_N=5.2\pm0.7 $, after calculating the absolute V magnitude ($M_{\rm V} =-21.5\pm0.14$) by means of the total $V_0$ magnitude obtained from NED and the adopted distance modulus. 
We can see that the specific frequency of the GCS of NGC\,3613 falls within the typical range expected for  early-type galaxies with similar  luminosity \citep{brodie2006,peng2008,georgiev2010,harris2013}. According to the model of GC formation presented by \citet{kruijssen2015}, where they use the definition of specific frequency normalized by host-galaxy stellar mass, the way GCs form from the interstellar medium in  discs and the subsequent disruption they suffer are the main physical processes shaping the behaviour of the specific frequency with respect to galaxy stellar mass.   

\section{Surface photometry of NGC~3613}
\label{sec:surface}
\begin{figure}
	\includegraphics[width=\columnwidth]{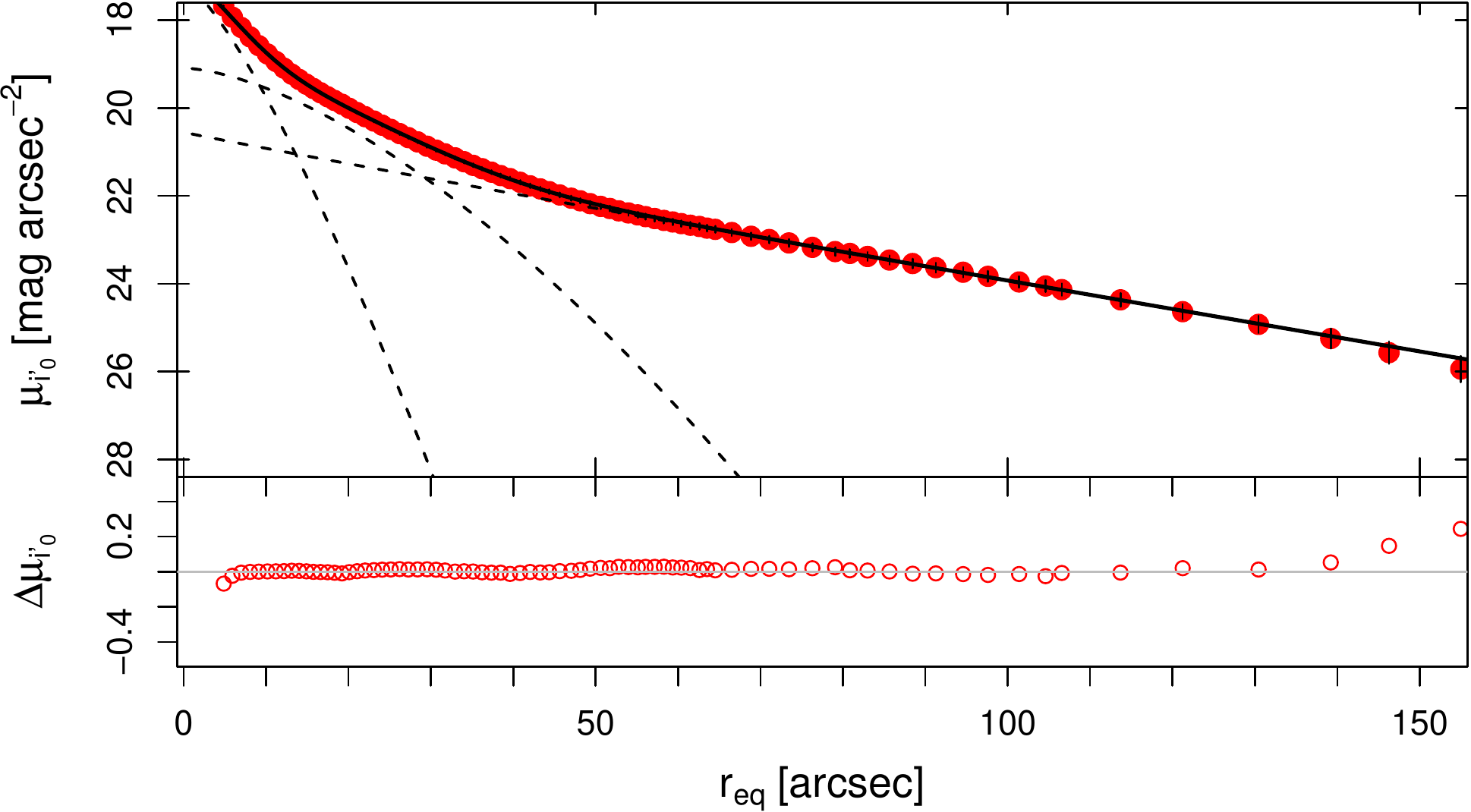}
    \caption{Surface-brightness profile of NGC\,3613 in the i'-band (top panel) and resulting residuals (bottom panel). Dashed and solid lines show the three S\'ersic models and their sum, respectively. The bottom panel represents the fit residuals. 
    }
    \label{fig:perfil}
\end{figure}

Fig.\,\ref{fig:perfil} (top panel) shows the surface-brightness profile of NGC\,3613 in the i'-band (reddening-corrected surface brightness $\mu_{i0}$ versus equivalent radius $r_{eq}$), obtained with \textsc{Iraf} through the \textmd{ELLIPSE} task. We used S\'ersic models to fit the galaxy profile and the best fit was provided by the addition of three components, as all fits with less components led to systematic residuals. 
The expression for each S\'ersic model is:

\begin{equation}
\mu(r_{eq})=\mu_{0}+1.0857\left(\frac{r_{eq}}{r_{0}}\right)^{1/n},
\end{equation}

where $\mu$ is the surface-brightness (in units of mag\,${\rm arcsec^{-2}}$), $\mu_{0}$ is the central surface-brightness, $r_{0}$ is a scale parameter and $n$ is the S\'ersic shape index (where n=1 corresponds to an exponential profile and n=4 to a de Vaucouleurs profile). The resulting residuals 
are shown in Fig.\,\ref{fig:perfil} (bottom panel). 
The parameters for the three fitted components are presented in Table~\ref{tab:paramperfil}. 
We have also included the respective effective radii, according to the relation 
$ r_{eff}=b_{n}r_{0}\, $,
where $b_n$ is a function of the n index, that may be estimated with the expression given by \citet{ciotti1991}.

\begin{table}
\begin{center}
\begin{tabular}{|c|c|c|c|}
\hline
   Parameter/Component & Inner & Intermediate & Outer \\
\hline \hline
$\mu_{0} [mag\,arcsec^{-2}]$ & $17.2\pm0.1$ & $19.1\pm0.1$ & $20.6\pm0.1$\\
\hline
$r_{0} [arcsec]$ & $5.2\pm0.2$ & $17.2\pm0.6$ & $30.8\pm1.9$\\
\hline
$n$ & $1.32\pm0.04$ & $1.58\pm0.04$ & $0.96\pm0.03$\\
\hline
$r_{\rm eff} [kpc]$ & $0.9$ & $2.4$ & $8.0$\\
\hline
 \end{tabular}
      \caption{Parameters of the three S\'ersic components fitted to the observed galaxy profile, i'-band.}
         \label{tab:paramperfil}
 \end{center}
\end{table}

The fitting parameters of our intermediate and outer components are in agreement within uncertainties with those of the bulge and exponential disk obtained by \citet{krajnovic2013} (ATLAS$^{3D}$ project) through a two-component fit. In particular, they point out that the median Sérsic index of the bulge is $n = 1.7$ for galaxies classified as fast rotators, i.e.  close to our value for NGC\,3613 ($n = 1.6)$.

The presence of three components in massive E galaxies, like our present target, has already been pointed out by several authors. For instance, \citet{huang2013} present a study of nearby Es from the Carnegie-Irvine Galaxy Survey and show that two-dimensional surface-brightness distributions of most of them, can be described by a compact core as inner component, an intermediate component as main body, and an outer envelope. For a sample close to 100 galaxies, they obtain S\'ersic index $n \approx 1-2$ for the components, in agreement with the values obtained for NGC\,3613 though we perform a one-dimensional analysis. Multi-components in this type of galaxies \citep[][and references therein]{huang2013,huang2013b,oh2017} are  understood as the consequence of a two-phase formation scenario. At high redshift (z $\geq 3$), the evolution is dominated by in-situ star formation owing to highly dissipative processes, from which the inner substructure of the galaxies derive. On the other side, the outer extended envelopes were built-up during a later phase, mainly dominated by accretion through `dry' minor mergers.

\begin{figure}
	\includegraphics[width=\columnwidth]{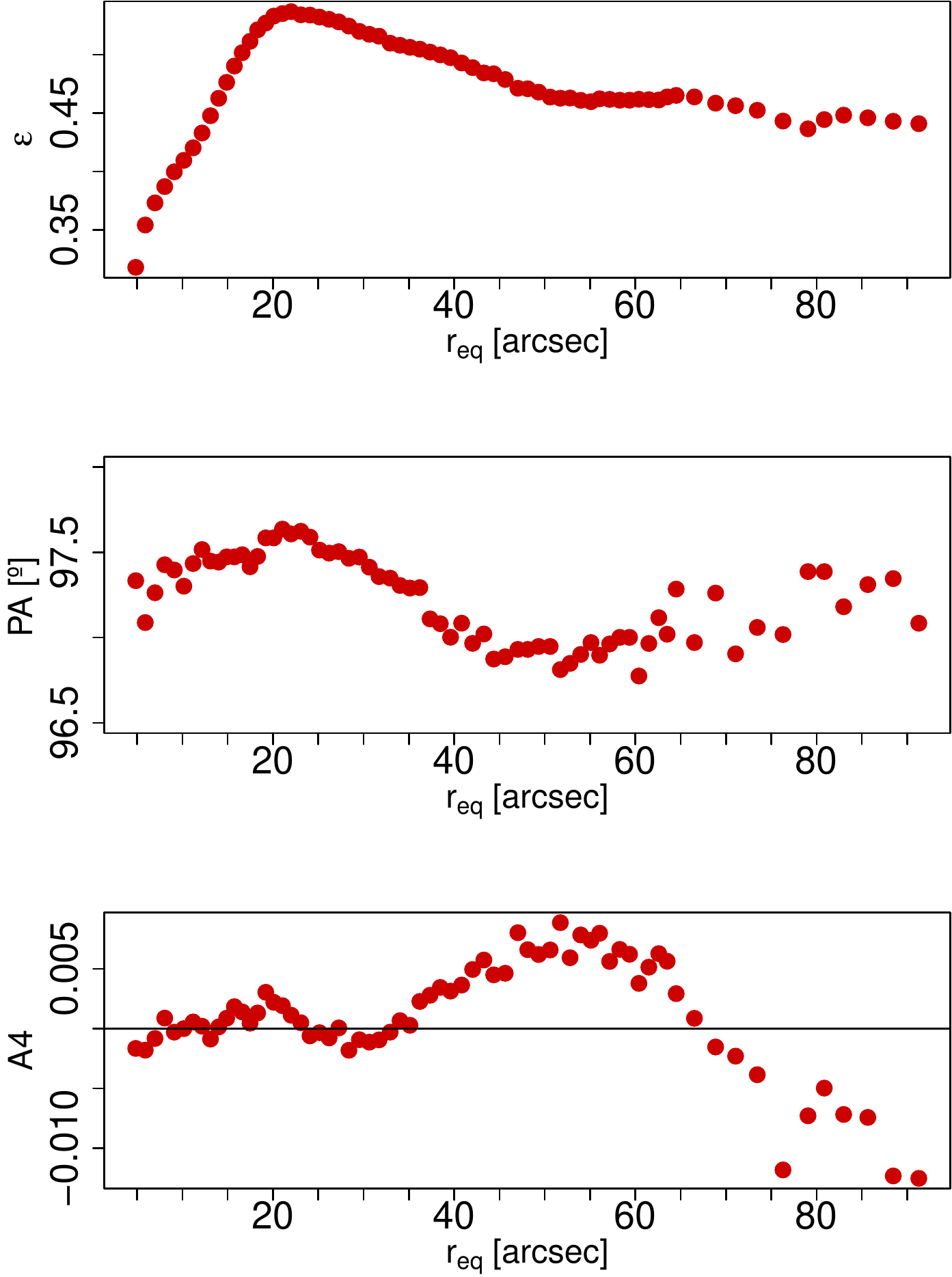}
    \caption{Parameters of the isophotes fitted with \textmd{ELLIPSE} to the NGC\,3613 surface-brightness distribution, as a function of $r_{eq}$. From top to bottom panels: ellipticity $\epsilon$, position angle PA, and $A4$ Fourier coefficient, respectively.
    The horizontal line in the bottom panel corresponds to  $A4=0$.
    }
    \label{fig:param}
\end{figure}

\begin{figure}
	\includegraphics[width=\columnwidth]{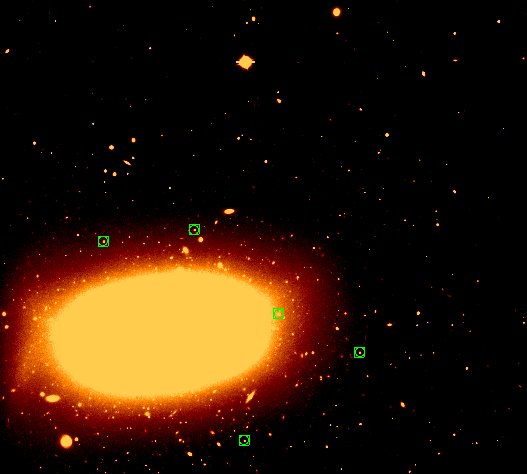}
    \caption{GMOS final image containing NGC\,3613  (i'-band). UCD candidates are highlighted with green  squares. North is up, east to the left.}
    \label{fig:original}
\end{figure}

Fig.\,\ref{fig:param} shows the parameters of the isophotes obtained with \textmd{ELLIPSE} for NGC\,3613, as a function of $r_{eq}$. That is, ellipticity $\epsilon$ (top panel), position angle PA measured positive from N to E (middle panel), and A4 Fourier coefficient which represents disky and boxy isophotes for $A4 > 0$ and $A4<0$, respectively (bottom panel). The values of $\epsilon$ are mostly higher than 0.4, which is typical of fast rotators as NGC\,3613 \citep{cappellari2011b}. 
Changes in the isophotal parameters, 
at $r_{eq} \sim 20$ and $\sim 55$\,arcsec, agree with the dominance of different components in the brightness profile.
Fig.\,\ref{fig:original} shows the final combined GMOS image (i'-band) of NGC\,3613, 
where the boxy shape of the outer isophotes is evident ($A4<0$). Five UCD candidates have been identified with squares in the surroundings. Globally, $\epsilon$ and PA agree with those given by \citep{krajnovic2011} in the context of the ATLAS$^{3D}$ project. 

\begin{figure}
	\includegraphics[width=\columnwidth]{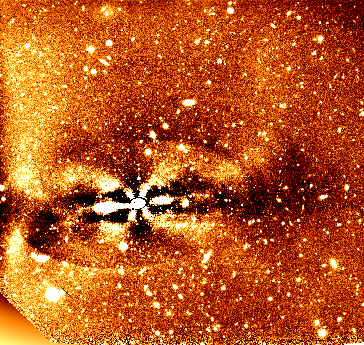}
    \caption{Residual GMOS image of NGC\,3613 (i'-band, FOV: $4.3 \times 4.3$\,arcmin, scale=0.146\,arcsec pixel$^{-1}$)}, obtained by subtracting a smooth model of the galaxy light.  
    Plumes in the outskirts and inner substructure are discernible. North is up, east to the left.
    \label{fig:restado}
\end{figure}

Fig.\,\ref{fig:restado} shows the GMOS image obtained by subtracting, from the original image, a smoothed model of the surface-brightness distribution of the galaxy, performed with \textmd{ELLIPSE} and \textmd{BMODEL}. In this residual image, there is an observable substructure at a low surface-brightness level. 
There is a plume towards the left side of the galaxy, pointing to the south, that is detectable in the original image (Fig.\,\ref{fig:original}) so that it cannot be a spurious residual of the image processing. Another plume is present on the opposite side, pointing to the north too. A bright x-shape residual located in the central region 
 may be connected to these plumes. An inner stellar disk is aligned with the major axis of the galaxy isophotes \citep[see also][]{ebneter1988}. All this lying substructure can be understood as another indication of the multi-components identified in the galaxy, related to the formation history, where the plumes may be tidal remainings of past accretions \citep[e.g.][]{barnes1992,hernquist1992}. On the other hand, we find no clear evidence of interaction with NGC\,3610.  

\section{Discussion}
\label{sec:discu}
\subsection{Relation between GC subpopulations and the host galaxy}

In many early-type galaxies, a close relationship has been observed between its stellar component and the red GC subpopulation, detected in the kinematics 
\citep[e.g.][]{pota2013}, in their radial projected distributions \citep[e.g.][]{youkyung2019} as well as the shape (measured by $\epsilon$) of the red clusters and the stellar light distribution \citep[e.g.][]{park2013}. 
In the case of NGC\,3613, we do not have enough radial coverage to determine colour gradients but can analyse the trend of mean colours for blue and red GCs, at three different radial ranges (Table~\ref{tab:paramgmm}). The blue peak of the inner radial range is redder than those of the intermediate and outer ranges, while the red peak does not present any clear variation with radius. That the blue peak gets bluer with increasing radius is in agreement with massive elliptical galaxies located at the centre of clusters (e.g., \citealt{bassino2006b} in Fornax, \citealt{caso2017} in Antlia), although  NGC\,3613 is considered as just the central galaxy of a group.

Regarding the red GC population, 
we noticed that the PA of the two over-densities detected in their projected azimuthal distribution 
(i.e., PA $\sim 110^{\circ}$ and $\sim 290^{\circ}$) correspond, as expected, to the orientation of the major-axis of the galaxy elliptical contours. 
In addition, the ellipticity of the projected distribution of the red clusters resulted $\epsilon = 0.37$. 
We calculated a mean $\epsilon$ and PA for the host galaxy isophotes with semi-axis between 48 $ < R_g < $ 102 (i.e. the same radial range 
used for the GC azimuthal distribution), resulting $<\epsilon> = 0.47$ ($\sigma = 0.017$) and $<PA> = 97^{\circ}$ ($\sigma = 0.14$). The shape parameters of the light distribution are very similar to those of the projected red GC distribution, while there is no obvious relation to the blue GCs. 

Both effects can be related to the formation history of the galaxy as it is generally accepted that most massive early-type galaxies in the local Universe form in two-phases \citep{forbes2011,vandokkum2015}.

\subsection{Ultra-compact dwarf candidates}
Five UCD candidates have been detected in the colour-magnitude diagram of point-sources, shown in Fig.\,\ref{fig:dcm} with  empty circles. This small sample, according to our photometry, shows colours within the range corresponding to GCs but their $i'$ magnitudes are brighter than expected for a GC (assuming a limit at $ i'_0 = 20.8 $, as explained in Section\,4.1). 

The positions of these UCD candidates are also identified on our GMOS image (Fig.\,\ref{fig:original}), where they appear surrounding NGC\,3613 at galactocentric radius between 66 and 121\,arcsec, i.e. well within the radial range covered by the GC candidates. Their colours are in the range $0.85<(g-i)_{0}<1.02$ and their absolute magnitudes $-11.8<M_i<-11.5$ according to adopted distance. If we compare to the $M_i$ versus $(g-i)_{0}$ colour-magnitude diagram presented by \citet[][their fig. 5]{brodie2011} for the sample of M\,87 UCDs, our candidates fall in the same locus as the M\,87 ones.

We plan to obtain spectra of these UCD candidates in the near future, in order to confirm membership with radial velocities and analyse physical properties like metallicity, age, stellar populations, etc.

\section{Summary and conclusions}
\label{sec:conclu}

We present the first photometric study of the GCS of the bright elliptical galaxy NGC\,3613, that is located at the centre of a galaxy group but has an intrinsic brightness typical of a brightest cluster galaxy. On the basis of $g'$, $r'$, $i'$ Gemini/GMOS images, not only the properties of the GCS but also the surface-photometry of the host galaxy were investigated. In addition, its distance was confirmed by means of the GC luminosity function and five new UCD candidates were discovered. The principal results are summarised here:

\begin{itemize}
    \item The GC colour distribution is bimodal, considering the whole sample or three different radial ranges. The mean colour of the blue GCs gets slightly bluer for increasing radius, which is understood as a hint that these metal-poor clusters may have been accreted with satellite galaxies. 
    
    \item Regarding the blue GC subpopulation, they follow a colour-magnitude relation in the sense that brighter clusters get redder, i.e. the so-called blue tilt, for whose interpretation several scenarios have been proposed. No equivalent relation is present in the  red GC subpopulation.
    
    \item Regarding the red GC subpopulation, its spatial, radial and azimuthal projected distributions show that they are more concentrated towards the host galaxy and trace closely the shape of the galaxy light isophotes. Thus, these effects point to a common origin of the galaxy stellar component and the majority of metal-rich GCs. The blue GC subpopulation presents a mostly uniform and more extended projected distribution. 
    
    \item By means of the turn-over of the GC luminosity function, we obtain a distance of $29.8 \pm 2.8$\,Mpc, in agreement within  uncertainties with the initially adopted value of 30.1\,Mpc  \citep{tully2013}. The total GC population is estimated in $N_{\rm tot}= 2075\pm133$ GCs and the specific frequency  $S_N=5.2\pm0.7$. Both values are typical for GCSs in host galaxies of similar luminosity than NGC\,3613.   
    
    \item There is a noticeable substructure in the surface-brightness distribution of NGC\,3613, detected in the original and residual images. It may be a sign of past tidal interactions but cannot be clearly related to any interplay with its neighbour, the merger remnant NGC\,3610. We also find no evidence of such interaction in the GC projected distributions. 
    
    \item We find a sample of five new UCD candidates in the outskirts of NGC\,3613, brighter than the regular GCs but within the same colour range. We plan to continue studying them with spectroscopy in the near future. 
    
\end{itemize}

\section*{Acknowledgements}
We thank the constructive comments of the referee, which helped to improve this paper. This work was funded with grants from Consejo Nacional de Investigaciones   
Cient\'{\i}ficas y T\'ecnicas de la Rep\'ublica Argentina, Agencia Nacional de Promoci\'on Cient\'{\i}fica y Tecnol\'ogica, and Universidad Nacional de La Plata, Argentina. \\
Based on observations obtained at the Gemini Observatory (programme GN2013A-Q-42, PI: J.P. Caso), which is operated by the Association of Universities for Research in Astronomy, Inc., under a cooperative agreement with the NSF on behalf of the Gemini partnership: the National Science Foundation (United States), the National Research Council (Canada), CONICYT (Chile), the Australian Research Council (Australia), Minist\'{e}rio da Ci\^{e}ncia, Tecnologia e Inova\c{c}\~{a}o (Brazil) and Ministerio de Ciencia, Tecnolog\'{i}a e Innovaci\'{o}n Productiva (Argentina). This research has made use of the NASA/IPAC Extragalactic Database (NED) which 
is operated by the Jet Propulsion Laboratory, California Institute of Technology, under contract with the National Aeronautics and Space Administration. 




\bibliographystyle{mnras}
\bibliography{biblio} 

\bsp	
\label{lastpage}
\end{document}